\documentclass[runningheads]{llncs}
\usepackage[T1]{fontenc} 
\usepackage{graphicx}

\usepackage{epstopdf}
\usepackage{color}
\usepackage[table]{xcolor}
\usepackage{makecell}
\usepackage{cite}
\usepackage{multirow}
\usepackage{amsfonts,amssymb,amsmath}
\usepackage{longtable}
\usepackage{booktabs}
\usepackage{xcolor}
\usepackage{floatrow}
\usepackage{subcaption} 
\newfloatcommand{capbtabbox}{table}[][\FBwidth]
\usepackage{marvosym}
 \usepackage[colorlinks]{hyperref}
\begin{document}
\title{Diff-UNet: A Diffusion Embedded Network\\
for Volumetric Segmentation}

\author{Zhaohu Xing\inst{1} \and
Liang Wan\inst{1}\and
Huazhu Fu\inst{2} \and
Guang Yang\inst{3} \and
Lei Zhu\inst{4,5} \textsuperscript{(\Letter)}
}

\institute{Medical College of Tianjin University, 
Tianjin, China \\
\email{xingzhaohu@tju.edu.cn} \and
 Institute of High Performance Computing, A*STAR  \and
Imperial College London \and
The Hong Kong University of Science and Technology (Guangzhou), Guangzhou, China \and
The Hong Kong University of Science and Technology, Hong Kong, China \
}
\maketitle              
\begin{abstract}





In recent years, Denoising Diffusion Models have demonstrated remarkable success in generating semantically valuable pixel-wise representations for image generative modeling. In this study, we propose a novel end-to-end framework, called Diff-UNet, for medical volumetric segmentation. Our approach integrates the diffusion model into a standard U-shaped architecture to extract semantic information from the input volume effectively, resulting in excellent pixel-level representations for medical volumetric segmentation.
To enhance the robustness of the diffusion model's prediction results, we also introduce a Step-Uncertainty based Fusion (SUF) module during inference to combine the outputs of the diffusion models at each step. We evaluate our method on three datasets, including multimodal brain tumors in MRI, liver tumors, and multi-organ CT volumes, and demonstrate that Diff-UNet outperforms other state-of-the-art methods significantly. Our experimental results also indicate the universality and effectiveness of the proposed model.
The proposed framework has the potential to facilitate the accurate diagnosis and treatment of medical conditions by enabling more precise segmentation of anatomical structures. 
The codes of Diff-UNet are available at 
\href{https://github.com/ge-xing/Diff-UNet}{https://github.com/ge-xing/Diff-UNet}. 

\keywords{Diffusion model  \and Medical Segmentation \and Volumetric Data.}
\end{abstract}

\section{Introduction}

Medical volumetric segmentation is a critical task for medical image analysis~\cite{litjens2017survey,khan2014survey,shamshad2201transformers}, involving the identification of lesion areas in high-dimensional medical image datasets on a pixel-by-pixel basis. More accurate segmentation results can provide valuable information to doctors, assisting them in diagnosing diseases.
Conventional 3D medical segmentation algorithms typically employ an encoder-decoder structure~\cite{ronneberger2015u,7785132,3d_unet} and incorporate skip-connections to enable the decoder to reuse features extracted by the encoder. Many current 3D medical image segmentation algorithms designed for model structures achieve promising segmentation results. For instance, SegResNet~\cite{myronenko20183d} uses variational Auto-Encoder~\cite{kingma2013auto} to add reconstruction branches, improving the feature extraction capability of the model. However, since its structure is based on a convolutional neural network, it may not be able to extract global features effectively.

Recenlty, the Transformer structure has gained popularity in modeling global features due to its global self-attention mechanism~\cite{vaswani2017attention,Shamshad2022}. TransBTS~\cite{wang2021transbts} leverages 3D-CNN to extract local spatial features and then applies the transformer to model global dependencies in high-level features. UNETR~\cite{hatamizadeh2022unetr} utilizes ViT~\cite{dosovitskiy2020image} as an encoder to model global features directly and outputs segmentation results using a CNN-based decoder with skip connections. However, the above methods are limited in their ability to extract multi-scale features due to the computational complexity of the Transformer structure. 
SwinUNETR~\cite{hatamizadeh2022swin} leverages Swin-Transformer~\cite{liu2021swin} as an encoder to extract multi-scale features and employs a CNN-based decoder to generate the output, achieving state-of-the-art medical image segmentation results.


Denoising diffusion models~\cite{ho2020denoising,song2020denoising,nichol2021improved} have shown significant success in various generative tasks, including medical image segmentation. For instance, MedSegDiff~\cite{wu2022medsegdiff} achieves 2D medical image segmentation by segmenting Denoising-UNet, and interacting with inter-structural information through Fourier transform. Wolleb et al.~\cite{wolleb2022diffusion} employ the diffusion model to solve the 2D medical image segmentation problem and improve the robustness of the segmentation results by fusing the output results of each diffusion step using a summation manner during testing. However, these methods are limited to 2D segmentation, and the diffusion model cannot generate multi-label segmentation directly.


Compared to traditional segmentation methods, the Diffusion model introduces noise at the input and iteratively predicts the segmentation label map, both of which can improve the robustness of the Diffusion model's prediction. To exploit the Diffusion model's potential, we propose a generic Diffusion-based end-to-end 3D medical image segmentation algorithm, called \textbf{Diff-UNet}, to solve the high-dimensional medical image segmentation problem.
However, the conventional diffusion model can only solve the binary segmentation problem. To segment multiple class, we design a Label Embedding operation that converts the segmentation label map into one-hot labels. This enables Diff-UNet to segment multiple targets simultaneously.
To extract semantic information from the input volume, we design a Denoising module that contains a Denoising-UNet and an independent Feature Encoder to learn the denoising process. This module outputs a clear segmentation label map from a noisy label map.
Finally, we design a Step-Uncertainty based Fusion (SUF) module that fuses multiple predictions from the Denoising module to obtain more robust segmentation results during the testing phase.
Extensive experiments on the BraTS2020 multimodal brain tumor segmentation dataset~\cite{menze2014multimodal,bakas2018identifying}, BTCV multi-organ segmentation dataset~\cite{BTCV}, and MSD Liver and Liver tumor segmentation dataset~\cite{antonelli2022medical} demonstrate that our method significantly outperforms state-of-the-art approaches.
\footnote{We will release our code after acceptance.}

\section{Method}

\begin{figure}[!t]
\includegraphics[width=0.9\textwidth]{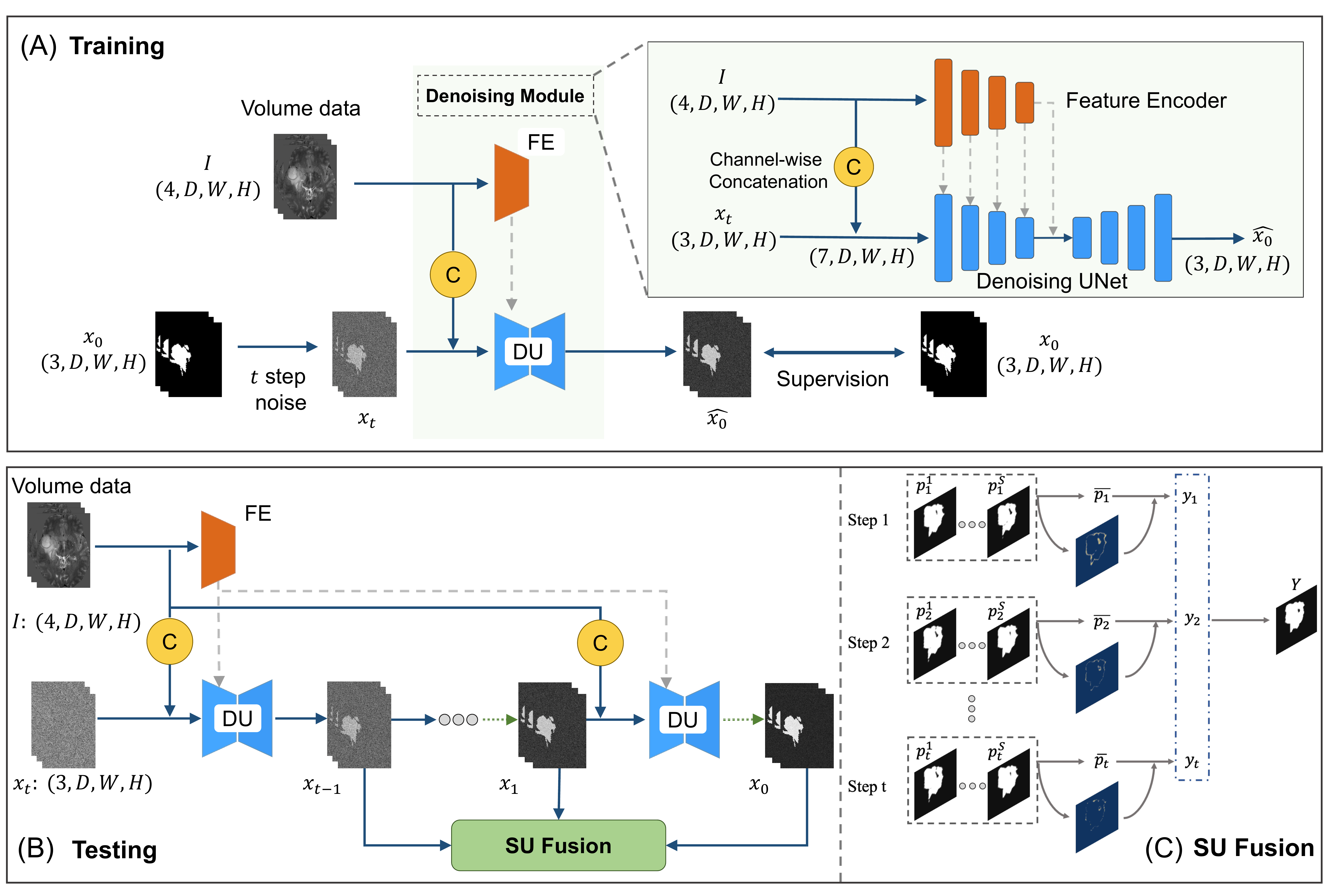}
\caption{The overview of proposed Diff-UNet. (A) is the training phase of Diff-UNet to learn a denoising function by the Denoising Module. (B) is the testing phase to generate segmentation results by an iterative way. (C) is the computation process of our SUF module. 
} \label{fig1}
\end{figure}

Fig.~\ref{fig1} shows the training stage, the testing stage, and two sub-modules of our proposed Diff-UNet: Denoising module and Step-Uncertainty based fusion module (SUF).
Unlike conventional medical image segmentation methods that directly input volume data to predict the corresponding segmentation labelmap, the diffusion model learns the denoising process. The diffusion model takes the volumetric image and the segmentation labelmap with noise as input and learns to remove the noise to generate clear segmentation results.

\subsection{Label Embedding}
One-hot encoding is usually used to convert multi-categorical tags into multiple bicategorical tags.
A one-hot vector $v$ is a binary vector of length $c$ where only a single entry can be one, all others must be zero. For example, assuming that there are 3 segmentation targets, we convert the segmentation labels $(0,1,2)$ by one-hot encoding to $((0,0,1),(0,1,0),(1,0,0))$.
The traditional Diffusion model only generates continuous data, which cannot predict the multi-target labels. Therefore, we first convert single-channel labelmap with size $D \times W \times H$ to multi-channel labels: ${x}_0 \in \mathbb{R}^{N \times D \times W \times H}$ by the one-hot encoding, where $N$ is the number of label, $(D, W, H)$ is the spatial resolutions of volumetric medical image.
Then, we add successive $t$ step noise $\epsilon$ for the converted multi-channel labels, called the diffusion forward process.
\begin{equation}
    \label{Eq:x_t}
    \mathbf{x}_{t}=\sqrt{\bar{\alpha}_{t}} \mathbf{x}_{0}+\sqrt{1-\bar{\alpha}_{t}} \boldsymbol{\epsilon} \ .
\end{equation}
After getting the labelmap $x_t$ with $t$ step noise, our target is to predict the clear labelmap $x_0$ based on $x_t$ and the raw volume data by the Denoising Module.
\subsection{Denoising Module}
As shown in Fig.~\ref{fig1} (A), the Denoising Module consisting of a Feature Encoder (FE) and a Denoising-UNet (DU) is a main part of Diff-UNet. 
The Denoising-UNet also contains two parts, an encoder, and a decoder. First, given the volume data $I \in \mathbb{R}^{M \times D \times W \times H}$, where $M$ is the number of modal images, $I$ and the noisy one-hot label $x_t$ are concatenated channel-wise into DU's encoder to obtain the multi-scale feature $\hat{I}_f:[\mathbb{R}^{if \times \frac{D}{i} \times \frac{W}{i} \times \frac{H}{i}}]_{i=1}^{16}$, where $f$ is the feature size and $i$ is the scale. Meanwhile, to better introduce the raw volumetric image features, we extract the multi-scale features $\tilde{I}_f$ of the volume data through a feature encoder which has the same size with the DU's encoder. Since $\tilde{I}_f$ and $\hat{I}_f$ contain the same number and size of features, we sum the features of the corresponding scales to obtain the fused features.
After that, we input the fused multi-scale features to the decoder of DU network to obtain the prediction result $\hat{x_0} \in \mathbb{R}^{N \times D \times W \times H}$:
\begin{equation}
    \hat{x_0} = \mathrm{DU}(\mathrm{cat}(I, x_t), t, \tilde{I}_f).
\end{equation}
Classical diffusion is trained using $\mathcal{L}_2$ denoising loss. In this task, we model the medical image segmentation task as a discrete data generation problem and directly predict $x_0$ instead of noisy $\epsilon$. Diff-UNet is trained by combing Dice Loss, BCE Loss, and MSE Loss, and thus the total loss $\mathcal{L}_{total}$ of our Diff-UNet is: 
\begin{equation}
    \mathcal{L}_{total} = \mathcal{L}_{dice}(\hat{x_0}, x_0) + \mathcal{L}_{bce}(\hat{x_0}, x_0) + \mathcal{L}_{mse}(\hat{x_0}, x_0).
\end{equation}

\subsection{Step-Uncertainty based Fusion}

The diffusion model iterates $t$ times in the testing phase by the Denoising Diffusion Implicit Models (DDIM) method. In a conventional generation task, the last prediction is taken as the final generation result, while each iteration of Diff-UNet generates a segmentation map. As the prediction time step increases, the more accurate the prediction result is and the lower the prediction uncertainty is. Therefore, to improve the segmentation robustness of Diff-UNet model, we fuse the output based on the number of prediction steps and uncertainty.

The way we compute the uncertainty is similar to the Monte Carlo Dropout (MC Dropout)~\cite{gal2016dropout}, which activates the network's dropout layer, then performs $S$ forward passes to estimate the uncertainty map. On the other hand, Diff-UNet initializes a random noise $x_t$ in the testing phase (Fig.~\ref{fig1} (B)), so it can introduce randomness into the network without activating the dropout layer. Like Monte Carlo dropout, Diffusion's testing process consists of $t$ steps, and each step predicts $S$ outputs, which are used to calculate the uncertainty. The formula is as:
\begin{equation}
\label{Eq:u_i}
    u_i = -\bar{p}_i log(\bar{p}_i) ,~ \text{where} ~\bar{p}_{i} = \frac{1}{S}\sum_{s=1}^{S}{p_i^s}.
\end{equation}
The fusion weights combining the number of prediction steps and uncertainty are calculated as $w_{i} = e^{\sigma(\frac{i}{scale}) \times (1 - u_{i})}$, where $\sigma$ is sigmoid funciton, $i$ denotes the current prediction step and $u$ is the uncertainty matrix.
We use $w$ to weight the prediction results of each step to obtain the final fusion result $Y$, which is used as the output of our network. Finally, $Y$ is given by: $Y = \sum_{i=1}^{t}{w_i \times \bar{p}_i}$.

\section{Experiments}


\noindent
\textbf{Implementation Details.} \ Our network is implemented in Pytorch and MONAI on 4$\times$ NVIDIA A100  GPUs. In the training phase, the loss function combines DICE loss, BCE loss, and MSE loss. We adopt an AdamW optimizer with a weight decay of 10-5. The warmup is set as 1/10 of the total number of epochs, and the learning rate is updated using the Cosine Annealing schedule. Each iteration randomly samples n patches (patch size is 96$\times$96$\times$96) for training.
Random flips, rotations, intensity scaling, and shifts are introduced for data augmentation.
In testing, we set the number of DDIM sampling steps as 10, and the size of each sample is 96$\times$96$\times$96. The sliding window overlap rate is 0.5 until the whole volume is predicted.

\subsection{Datasets and Evaluation Metrics}

To evaluate the volumetric segmentation performance of our method, we utilize three publicly available segmentation datasets, including BraTS2020~\cite{menze2014multimodal,bakas2018identifying}, MSD Liver~\cite{antonelli2022medical} dataset, and the abdominal multi-organ segmentation dataset BTCV~\cite{BTCV}. 
Moreover, the Dice score and 95\% Hausdorff Distance (HD95) are adopted for quantitative comparison.

\noindent
\textbf{BraTS2020 dataset} contains 369 aligned four-modality MRI data (i.e., T1, T1ce, T2, FLAIR) with expert segmentation masks (i.e., GD-enhancing tumor, peritumoral edema, and tumor core). 
Each modality has a 155×240×240 volume and all modality images have already been resampled and co-registered. 
The segmentation task aims to segment the whole tumor (WT), enhancing tumor (ET), and tumor core (TC) regions.
The splitting ratio for the training set, the validation set, and the test set is 0.7, 0.1, and 0.2.

\noindent
\textbf{MSD Liver dataset} has a total of 131 cases of 3D liver images with 1 modality and 2 segmentation targets (liver and liver tumor) for each 3D liver image. All data are resampled to the same space (2.0, 2.0, 2.0).
%
The  MSD liver dataset is divided into a training set, validation set, and test set according to the ratio of 0.7, 0.1, and 0.2.

\noindent
\textbf{BTCV dataset} consists of 30 cases of 3D abdominal
multi-organ images and each 3D image has 13 organ segmentation targets.
All data are resampled to the same space (2.0, 1.5, 1.5). 
Following TransUNet, 18 cases are used for training, and the remaining 12 cases are for testing.

\subsection{Comparison with SOTA Methods}

\begin{table*}[t]
    \centering
    \caption{Quantitative comparison on BraTS2020 dataset.}
    \label{tab:brats_result}
    \renewcommand\arraystretch{1}
    \setlength\tabcolsep{3pt}
    \resizebox{\textwidth}{!}{
    \begin{tabular}{c c c c c c c  c c  c c c c}
    \Xhline{1pt}
    \multirow{2}{*}{Methods} & \multicolumn{2}{c}{WT} &  & \multicolumn{2}{c}{TC} & &  \multicolumn{2}{c}{ET} &  & & \multicolumn{2}{c}{Average} \\
    \cline{2-3} \cline{5-6} \cline{8-9} \cline{12-13}
      & Dice$\uparrow$ & HD95$\downarrow$ & & Dice$\uparrow$ & HD95$\downarrow$ & & Dice$\uparrow$ & HD95$\downarrow$ & & & Dice$\uparrow$ & HD95$\downarrow$  \\
    \hline
    SwinUNETR~\cite{hatamizadeh2022swin} & 91.68 & 2.856 &  & 82.60 & 4.314 &  & 74.85 & 4.503 &  & & 83.04 & 3.891  \\
    UNETR~\cite{hatamizadeh2022unetr} & 90.15 & 4.305 &  & 81.26 & 5.740 &  & 73.23 & 4.643 &  & & 81.55 & 4.896\\
    TransBTS~\cite{wang2021transbts} & 91.06 & 3.360 &  & 83.60 & 2.986 &  & 74.03 & \textbf{3.403} &  & & 82.90 & 3.249 \\
    SegResNet~\cite{myronenko20183d} & 91.54 & {3.2275} &  & 83.61 & 3.769 &  & 73.04 & 3.486 &  & & 82.73 & 3.494 \\
    Attention-UNet~\cite{oktay2018attention} & 84.49 & 15.174 &  & 78.17 & 16.380 &  & 71.62 & 9.095 &  & & 78.09 & 13.549  \\
    ModelsGenesis~\cite{zhou2021models} & {91.98} & 2.799 &  & 84.31 & \textbf{2.836} &  & 73.84 & 4.333 &  & & 83.38 & \textbf{3.096}  \\
    Our Diff-UNet & \textbf{92.23} & \textbf{2.588} &  & \textbf{86.94} & {3.596} &  & \textbf{76.87} & 3.984 &  & &  \textbf{85.35} & {3.389}\\
    \Xhline{1pt}
    \end{tabular}
    }
\end{table*}

\begin{table*}[t]
    \centering
    \caption{Quantitative comparison on MSD Liver dataset.}
    \label{tab:liver_result}
    \renewcommand\arraystretch{1}
    \setlength\tabcolsep{5pt}
    \resizebox{\textwidth}{!}{
    \begin{tabular}{c c c c c c c  c c  c}

    \Xhline{1pt}

    \multirow{2}{*}{Methods} & \multicolumn{2}{c}{Liver} &  & \multicolumn{2}{c}{Tumor} & & \multicolumn{2}{c}{Average} \\
    \cline{2-3} \cline{5-6} \cline{8-9}
      & Dice$\uparrow$ & HD95$\downarrow$ & & Dice$\uparrow$ & HD95$\downarrow$& & Dice$\uparrow$ & HD95$\downarrow$  \\
    \hline

    SwinUNETR~\cite{hatamizadeh2022swin} & 95.47 & 0.392 &  & 49.94 & 20.906 & & 72.70 & 10.645  \\
    UNETR~\cite{hatamizadeh2022unetr} & 93.75 & 1.080 &  & 38.43 & 24.87 &  & 66.09 & 12.979\\
    TransBTS~\cite{wang2021transbts} & 95.11 & 0.403 &  & 44.99 & 17.463 &  & 70.05 & 8.933 \\
    SegResNet~\cite{myronenko20183d} & 95.30 & 0.418 &  & 46.39 & 19.424 & & 70.85 & 9.921 \\
    Attention-UNet~\cite{oktay2018attention} & 95.32 & 0.499 &  & 48.43 & 20.273 & & 71.88 & 10.386  \\
    ModelsGenesis~\cite{zhou2021models} & {95.04} & 0.934 &  & 50.04 & 31.146 & & 72.54 & {15.823}  \\
    Our Diff-UNet & \textbf{95.72} & \textbf{0.222} &  & \textbf{51.65} & \textbf{17.280} &  &  \textbf{73.69} & \textbf{8.751}\\
    \Xhline{1pt}
    \end{tabular}
    } 
\end{table*}

\begin{table*}[t]
    \centering
    \caption{Quantitative comparison on BTCV muti-organ sementation dataset.}
    \label{tab:btcv_segmentation}
    \renewcommand\arraystretch{1}
    \setlength\tabcolsep{3pt}
    \resizebox{\textwidth}{!}{
    \begin{tabular}{c c c c c c c c c c c c c c c }
    \Xhline{1pt}
    
    \multicolumn{2}{c}{Framework} & \multicolumn{2}{c}{} &  \multicolumn{2}{c}{Average} & \multirow{2}{*}{Aorta} & \multirow{2}{*}{Gallbladder} & \multirow{2}{*}{Kidney(L)} & \multirow{2}{*}{Kidney(R)} & \multirow{2}{*}{Liver} & \multirow{2}{*}{Pancreas} & \multirow{2}{*}{Spleen} & \multirow{2}{*}{Stomach} \\
    \cline{1-2} \cline{4-6}
    Encoder & Decoder & &  & Dice$\uparrow$ & HD95$\downarrow$ & \\
    \hline
    \multicolumn{2}{c}{VNet} & & & 68.81 & - & 75.34 & 51.87 & 77.10 & 80.75 & 87.84 & 40.05 & 80.56 & 56.98 \\
    \multicolumn{2}{c}{DARR} & & & 69.77 & - & 74.74 & 53.77 & 72.31 & 73.24 & 94.08 & 54.18 & \textbf{89.90} & 45.96 \\
    R50 & U-Net & & & 74.68 & 36.87 & 84.18 & 62.84 & 79.19 & 71.29 & 93.35 & 48.23 & 84.41 & 73.92 \\
    R50 & AttUNet & & & 75.57 & 36.97 & 55.92 & 63.91 & 79.20 & 71.71 & 93.56 & 49.37 & 87.19 & 74.95 \\
    \hline
    ViT & None & & & 61.50 & 39.61 & 44.38 & 39.59 & 67.46 & 62.94 & 89.21 & 43.14 & 75.45 & 69.78 \\
    ViT & CUP & & & 67.86 & 36.11 & 70.19 & 45.10 & 74.70 & 67.40 & 91.32 & 42.00 & 81.75 & 70.44 \\
    R50-Vit & CUP & & & 71.29 & 32.87 & 73.73 & 55.13 & 75.80 & 72.20 & 91.51 & 45.99 & 81.99 & 73.95 \\
    \multicolumn{2}{c}{TransUNet} & & & 77.48 & 31.69 & 87.23 & 63.13 & 81.87 & 77.02 & 94.08 & 55.86 & 85.02 & \textbf{75.62} \\
    \multicolumn{2}{c}{Our Diff-UNet} & & & \textbf{83.75} & \textbf{8.115} & \textbf{89.30} & \textbf{76.23} & \textbf{85.20} &  \textbf{84.73} & \textbf{95.90} & \textbf{74.25} & 89.75 & 74.65 \\
    \Xhline{1pt}
    \end{tabular}
    } 
\end{table*}

For the BraTS2020 and MSD liver datasets, we compare our Diff-UNet against state-of-the-art segmentation methods, including SwinUNETR, UNETR, TransBTS, SegResNet, Attention-UNet, and ModelsGenesis. For a fair comparison, all methods use publicly available implementations.
For the BTCV dataset, we follow same experimental setting of TransUNet to utilize the same training and testing dataset and compare with state-of-the-art methods.
 
\noindent
\textbf{BraTS2020. } 
Table \ref{tab:brats_result} reports the Dice and HD95 scores and the average scores of all methods on the three regions (WT, TC, ET) for the BraTS2020 dataset. Apparently, our proposed Diff-UNet method clearly outperforms compared state-of-the-art methods in terms of the Dice score for all three regions and their average. The average Dice of the three regions achieves 85.35\%, which has a improvement of 1.97\% than the second place ModelsGenesis. Although our average HD95 score on three regions achieved 3.3898, which takes the 3rd rank, it is slightly smaller than the top two HD95 results (i.e., 3.2499 and 3.0961).

\noindent
\textbf{MSD Liver. }
Table \ref{tab:liver_result} reports the Dice and HD95 performance of our proposed Diff-UNet and state-of-the-art methods on the MSD Liver dataset. Compared with other comparison methods, our Diff-UNet method has the larger Dice score and the smaller HD95 score on the Liver region, the tumor region and their average. 
Specifically, our Diff-UNet achieves a Dice score of 95.72\% and a HD95 score of 0.222 on Liver segmentation, and a Dice score of 51.65\% and a HD95 score of 17.280 on Liver tumor segmentation. And the average Dice and HD95 score on two regions are 73.69\% and 8.751. 

\noindent
\textbf{BTCV.}
Following the same experimental setup of TransUNet, we report the Dice scores for the eight abdominal organs and the average Dice and HD scores in Table~\ref{tab:btcv_segmentation} to compare our Diff-UNet and state-of-the-art segmentation methods. 
From Table~\ref{tab:btcv_segmentation}, we can find that our Diff-UNet has the best averaged Dice and HD95 scores on eight organs.
Specially, our Diff-UNet takes the 1st rank on the Dice score for six organs, and the averaged Dice and HD95 scores are 83.75\% and 8.115.
It indicates that our Diff-UNet can achieve a more accurate multi-organ segmentation performance than state-of-the-art methods in the BTCV dataset.\
Although our Diff-UNet takes the 2nd Dice rank on Spleen, and the 3rd Dice rank on Stomach, their Dice scores (89.75\% and 74.65\%) are slightly smaller than the best ones, which are 89.90\% for Spleen and 75.62\% for Stomach.

\begin{figure}[!t]
\includegraphics[width=\textwidth]{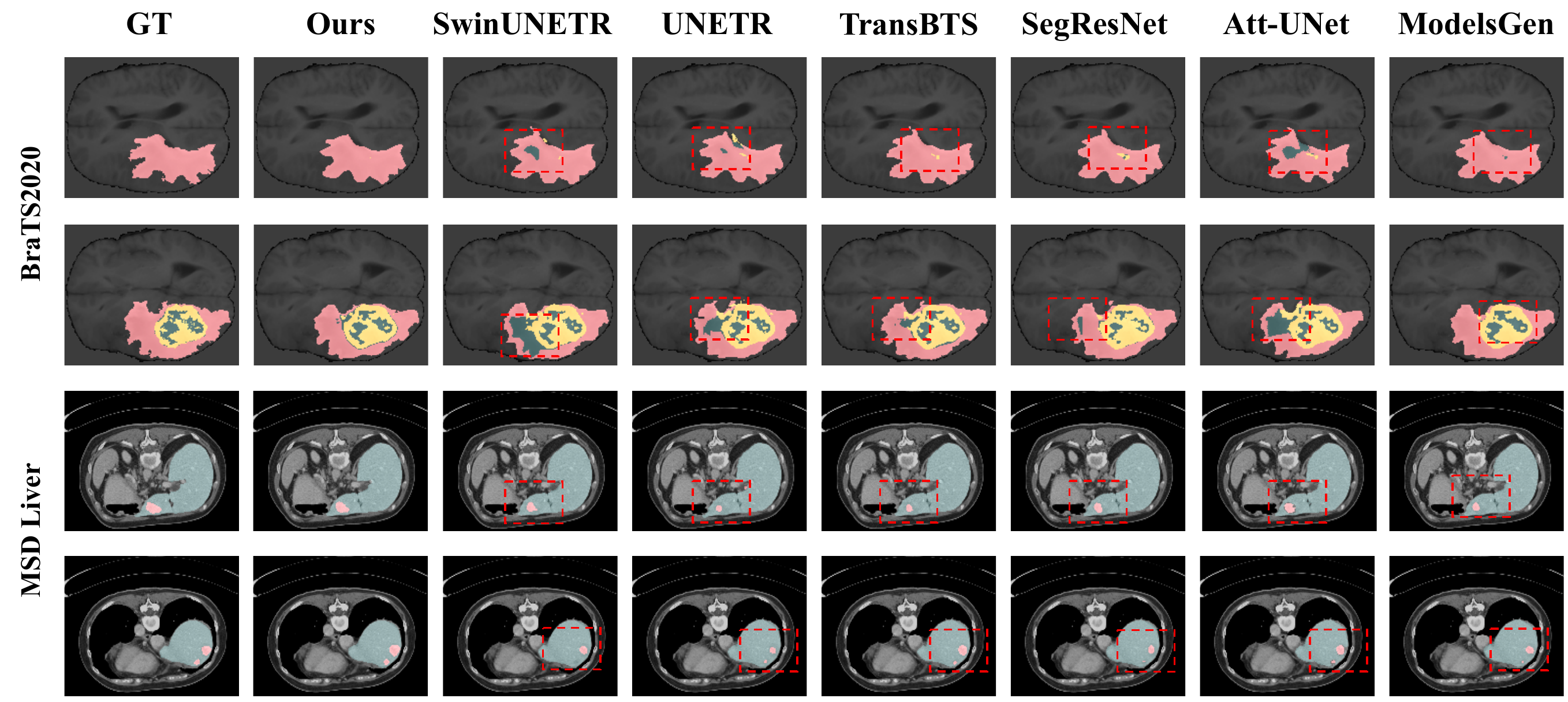}
\caption{The visual comparisons on segmentation results produced by our network and state-of-the-art methods on BraTS2020 and MSD Liver datasets. Apparently, our method has a more accurate segmentation performance and is consistent with the ground truth (denoted as ``GT'').} \label{fig:visual}
\end{figure}

\noindent
\textbf{Visual Comparisons.}
Fig.~\ref{fig:visual} visually compares the segmentation results produced by our Diff-UNet and SOTA methods on BraTS2020 and MSD Liver datasets. 
We do not run the comparison method on the BTCV dataset but use the result table in TransUNet directly, so we do not show the segmentation results on the BTCV dataset.
We can find that our Diff-UNet achieves more accurate segmentation results, especially on the tiny targets, while compared methods tend to miss some target regions, or include other non-target regions (see SwinUNETR at the 2nd row) in their segmentation results. 

\subsection{Ablation study}
\begin{table*}[tp]
\small 

\begin{floatrow}
\resizebox{0.58\textwidth}{!}{
\centering
\setlength\tabcolsep{3pt}
\renewcommand\arraystretch{1.0}
\ttabbox{\caption{Ablation study for different modules on BraTS2020 dataset. FE denotes the separate Feature Encoder. SF denotes a simple fusion. SUF denotes the Step-Uncertainty based Fusion module.}}{%
\vspace{-2mm}
\label{tab:ablation_module}
\begin{tabular}{c c c c c c} 
    \hline
    Module & & WT & TC & ET & Average \\
    \hline
    basic & &  91.62 & 85.02 & 75.10 & 83.91 \\
    basic+FE & & 91.52 & 85.85 & 75.59 & 84.32 \\
    basic+FE+SF & &  92.02 & 86.58 & 75.67 & 84.76 \\
    basic+FE+$\mathrm{SUF}$ (Ours) & & \textbf{92.23} & \textbf{86.94} & \textbf{76.87} & \textbf{85.35}\\
    \hline
    \end{tabular}
    }
}


\resizebox{0.42\textwidth}{!}{
\centering
\setlength\tabcolsep{3pt}
\renewcommand\arraystretch{1.0}
\begin{floatrow}
\ttabbox{\caption{Ablation study for the number (S) of predictions to compute the uncertainty in each DDIM step on BraTS2020 dataset. }
\vspace{-2mm}
\label{tab:ablation_S}}{%
    \begin{tabular}{c c c c c c} 
    \hline
    $S$ & & WT & TC & ET & Average \\
    \hline
    3 & &  92.19 & 86.18 & 76.82 & 85.06 \\
    4 (Ours) & & \textbf{92.23} & 86.94 & \textbf{76.87} & \textbf{85.35} \\
    5 & &  92.17 & \textbf{86.96} & 76.84 & 85.32 \\
    6 & & 92.22 & 86.92 & 76.84 & 85.33 \\
    \hline

    \end{tabular}
    }
\end{floatrow}
}
\vspace{-5mm}
\end{floatrow}
\end{table*}

\noindent
\textbf{Effectiveness of major modules.} We conduct ablation experiments on the BraTS2020 dataset to evaluate the role of different major modules (i.e., FE and SUF) involved in Diff-UNet and show their quantitative results in Table~\ref{tab:ablation_module}. 
From the quantitative results in Table~\ref{tab:ablation_module}, we can find that ``basic+FE'' has a larger averaged Dice on three regions (i.e., WT, TC, and ET) than ``basic'', which indicates that taking our FE as the image encoder can introduce more image information to the diffusion model, thereby improving the segmentation accuracy. 
Meanwhile, the superior Dice score of ``basic+FE+SF'' over ``basic+FE'' demonstrates that fusing the segmentation results predicted at each step of DDIM can further improve the segmentation accuracy of the diffusion model.
Moreover, our method has a superior Dice score over ``basic+FE+SF'', which shows that assigning different weights to integrate predictions at different DDIM steps can further enhance the segmentation performance of our method.

\noindent
\textbf{Setting S.} Moreover, we conduct an ablation study experiment to discuss how to set the value of $S$ (see Eq.~\ref{Eq:u_i}), which is the number of predictions to compute the uncertainty in each DDIM step. Here we consider different values for $S$, and they are 3, 4, 5, and 6 and show the corresponding results in Table~\ref{tab:ablation_S}.
Apparently, our method has the best averaged Dice score on three regions when $S=4$, and it has the largest Dice score of 92.23\% on WT, the second largest Dice score of 76.87\%, and the largest Dice score of 85.35\% on ET. 
Hence, we empirically set $S=4$ in our method.

\section{Conclusion}
In this paper, we propose the first 3D medical image segmentation method, named Diff-UNet, based on the diffusion model, which models medical image segmentation as a discrete data generation task. The proposed algorithm introduces a generic end-to-end 3D medical image segmentation approach, leveraging the advantages of the Diffusion model to improve segmentation robustness. Experimental results on different benchmark datasets demonstrate the superiority of our Diff-UNet over state-of-the-art approaches.
Overall, our work presents a significant contribution to the field of medical image segmentation, demonstrating the effectiveness of the Diffusion model in the 3D medical image segmentation task. The proposed method has the potential to facilitate more precise and accurate diagnosis and treatment of medical conditions, ultimately leading to improved patient outcomes.

\bibliographystyle{splncs04}
\bibliography{references}
%




\end{document}